**(FOR REVIEWERS)**

# Insights into glass formation and glass transition in supercooled liquids, by study of related phenomena in crystals.

C. Austen Angell
Department of Chemistry and Biochemistry,
Arizona State University, Tempe AZ 85287

**Abstract.**
We divide glass and viscous liquid sciences into two major research areas, the first dealing with how to avoid crystals and so access the viscous liquid state, and the second dealing with how liquids behave when no crystals form. We review some current efforts to elucidate each area, looking at strategies for vitrification of monatomic metals in the first, and the origin of the property "fragility" in the second. Essential here is the non-trivial behavior of the glassformer thermodynamics. We explore the findings on nonexponential relaxation and dynamic heterogeneities in viscous liquids, emphasizing the way in which direct excitation of the configurational modes has helped differentiate configurational from nonconfigurational contributions to the excess heat capacity. We then propose a scheme for understanding the relation between inorganic network and non-network glassformers which includes the anomalous case of water as an intermediate. In a final section we examine the additional insights to be gained by study of the ergodicity-breaking, glass-like, transitions that occur in disordering crystals. Here we highlight systems in which the background thermodynamics is understood because the ergodic behavior is a lambda transition. Water and the classical network glassformers appear to be attenuated versions of these.

## 1. INTRODUCTION

Glassy materials have served humanity from long before the beginning of recorded history. While the optical quality of obsidian glass knives and arrowheads was probably never an issue, the presence of tiny crystals in the glass lenses of early telescopes certainly was, and so the understanding of crystal formation and growth in glass became a matter of concern in the early days of glass science. Even earlier, the manner in which glassy materials solidified from the melt must have exercised the artisans who created glass objects of diverse and pleasing shapes by the skilful manipulation of the decreasing fluidity during cooling - and it was surely noticed how those glasses made with a large content of soda (sodium carbonate) solidified much more abruptly than those with little soda. And so there was surely, even in those early times, an appreciation of the property that glass scientists now call viscous liquid "fragility", and seek with much diligence to understand - so far with limited success.

Certainly, the first problem of glass science is to ensure that the cooling liquid does not become a mass of crystals, as thermodynamics alone dictates it should. Except for the case of atactic macromolecules, the lowest free energy states of non-quantum substances are always crystalline in character at low temperature. To the extent that this is true, then, glass formation is always a matter of arranging for the time-scale of crystal formation to be long relative to the rate at which we choose to cool the liquid. Since there are limits to how quickly we can cool liquids and since simple liquids seem to crystallize very quickly, there is a continuing challenge to find just how simple a liquid can be, and still be vitrifiable. Elements would seem to be rather simple, and so it may seem surprising that it is easy to obtain selenium as a glass. Its periodic table neighbor, sulfur, can also be vitrified if it is cooled quickly after heating in a special temperature range. But both of these cases depend on the ability of these elements to bind their atoms into complex polymeric forms.

## 2. AVOIDING CRYSTALLIZATION: GLASSES FROM SIMPLE LIQUIDS.

The simplest non-polymeric molecular liquid to vitrify easily is probably sulfur monochloride, which contains four atoms per molecule. The even simpler case of ozone satisfies an empirical rule, due to Cohen and Turnbull[1, 2], which predicts glass formation for liquids with melting points less than half their boiling points. However, ozone's reputation for instability and explosion has so far discouraged its evaluation as a glassformer.

Liquid metals, which are usually thought of as atomic in character, were long thought to be too simple to be vitrified, but then it was found that fast cooling of certain binary metallic alloys indeed resulted in glassy solids. Nowadays it is known[3] that the 1:1 binary alloy ZrCu need only be dripped onto a cold metal plate for it to form glassy discs - or poured into a 1mm diameter tubular graphite mold, to form glassy rods. This is a remarkably simple glassformer.

### Computer simulation studies.

Recently a new way of exploring glass formation (i.e. the exclusion of crystallization during slow cooling) has been developed. This method is not bounded by the table of elements, which gives it a major advantage. It also has the advantage of a redefinition of the meaning of the word "slow". We refer, here, to the study of liquid cooling by computer simulation. Specifically, we consider the use of molecular dynamics (MD) methods since in these studies the atoms or molecules behave very much the way they do in "real" liquids. Even the way in which heat is extracted to cool the liquid (by kinetic energy exchange with a colder reservoir) may be made the same as in the laboratory[4]. The difference is that even with the slowest cooling rates that current computation speeds permit, these simulated coolings are very fast by laboratory standards. In fact the feasible quenching rates still exceed what is possible in most laboratory quenches: there is, unfortunately, barely any overlap between the two.

However, the properties of liquids which, in the laboratory, bestow vitrifiability may be replicated in the MD experiments. For instance, it is found in laboratory studies using the fastest cooling rates available[5], that crystals may be avoided and thus glasses formed if, at the melting point (or liquidus temperature for multicomponent systems), the diffusion coefficient of the dominant species (the one that determines the viscosity) can be brought below the value $10^{-9}$ $m^2s^1$. This magnitude of diffusivitiy can be studied in simulations with relative ease. It is the investigation of melting points that has not been given much attention. Thus MD simulation provides an excellent tool for exploring the fundamental conditions needed for crystal avoidance during cooling. However it has, to date, been little used for this purpose.

MD has instead been used intensively to study the properties of simple liquids that do not crystallize on computation time scales (the subject of our second section). Foremost among such studies have been those investigating the "binary mixed Lennard-Jones" system (BMLJ)[4, 6]. In its initial incarnation, this binary atomic mixture was based on a binary metallic glassformer consisting mostly of nickel atoms together with some phosphorus, $Ni_3P$. This composition was known to be a "marginal" metallic glassformer from early laboratory studies[5, 7]. Initially, Weber and Stillinger[8] developed a binary atomic model with interaction potentials that closely reproduced the known structure (pair distribution functions) and liquid state behavior of $Ni_3P$. This then morphed into the so-called BMLJ system when its parameters were modified by Kob and Andersen[9] to more closely represent an interacting Lennard-Jones system. Despite very intensive study under many conditions, BMLJ in the Kob-Andersen version has never been known to crystallize in MD simulations. However, when the atomic ratio is changed from 3:1 to 1:1, it crystallizes rapidly[10]. What was achieved by Weber and Stillinger[8], Kob and Andersen[9], then, was the "chemical stabilization" of the liquid state relative to the free energy of any available crystals, thus ordaining the crystal nucleation kinetics to be slow.

The liquidus temperatures (in the binary phase diagram) of the BMLJ system have never been determined, and indeed they would be very difficult to determine in the glassforming composition range. Fortunately, however, the idea of stabilizing the liquid vis-à-vis the competing crystal structures, is not limited to binary systems. Such studies are in fact more fruitfully carried out on single component systems because in these cases the melting point can always be determined quite simply. The melting points of single component systems can, for instance, be changed by change of pressure. However a more interesting variable, available for MD simulated systems, is the fundamental interaction potential. It must be expected that the relative stabilities of liquid and crystal phases of the same atomic system will change with this interaction potential — and in simulations, unlike nature, there is no limit on the possible potential functions via which the atoms of the system can made to interact.

The idea of tuning potentials to induce (or modify) certain liquid properties of molecular and atomic systems, has been around for some time[11] [12], but had not been exploited for the specific purpose of determining the conditions for vitrification until rather recently. In 2006 Molinero et al[13] changed this by investigating potentials of the form developed by Stillinger and Weber[14] for the (quite successful) simulation of the element silicon (in which of course there is enormous technological interest, especially in the crystal growth from melt). Before summarizing Molinero et al's findings, an interesting and important feature of studies with the S-W silicon potential should be briefly revisited. This was the support that the behavior of its liquid gave to a challenging thermodynamic deduction by Spaepen and Turnbull[15] and Bagley and Chen[16] about laboratory silicon.

The latter authors had deduced that silicon contained, in its supercooled liquid state, a transition from the initial supercooled liquid to another amorphous state, which (by comparison with the diffusivity of known *crystal* diffusivity up to the deduced transition temperature[17]), had also to be a liquid state. The existence of such a liquid-liquid phase transition had been the theoretical prediction of a Russian scientist Aptekar[18] who (as colleague of Ponyatovsky[19]) had been involved in the study of other interesting isosymmetric phase transitions one of which will be discussed below. The liquid-liquid phase transition in silicon will play an important part in our developing understanding of factors that affect the rates at which crystals can form.

Returning to the main theme, Molinero et al examined what happened to the melting points, and the liquid state properties, of atomic systems as the parameter determining the strength of the "tetrahedrality" parameter **λ** in the three-body part of the S-W silicon potential was changed. They found that the melting point of the normal diamond cubic phase of Si decreased rapidly with weakening of the "tetrahedrality" of the potential and that the diffusivity of the liquid at the melting point systematically decreased (although the *isothermal* diffusivity increased). This continued until the melting point reached 50 % of its initial value, although the cohesive energy (determined by the two-body attractive potential) did not change. For λ values lower than 19, a new crystalline phase (body centered cubic, BCC) of higher coordination number became the more stable and this caused the melting point to increase. The T-λ phase diagram for this system is shown in Figure 1.

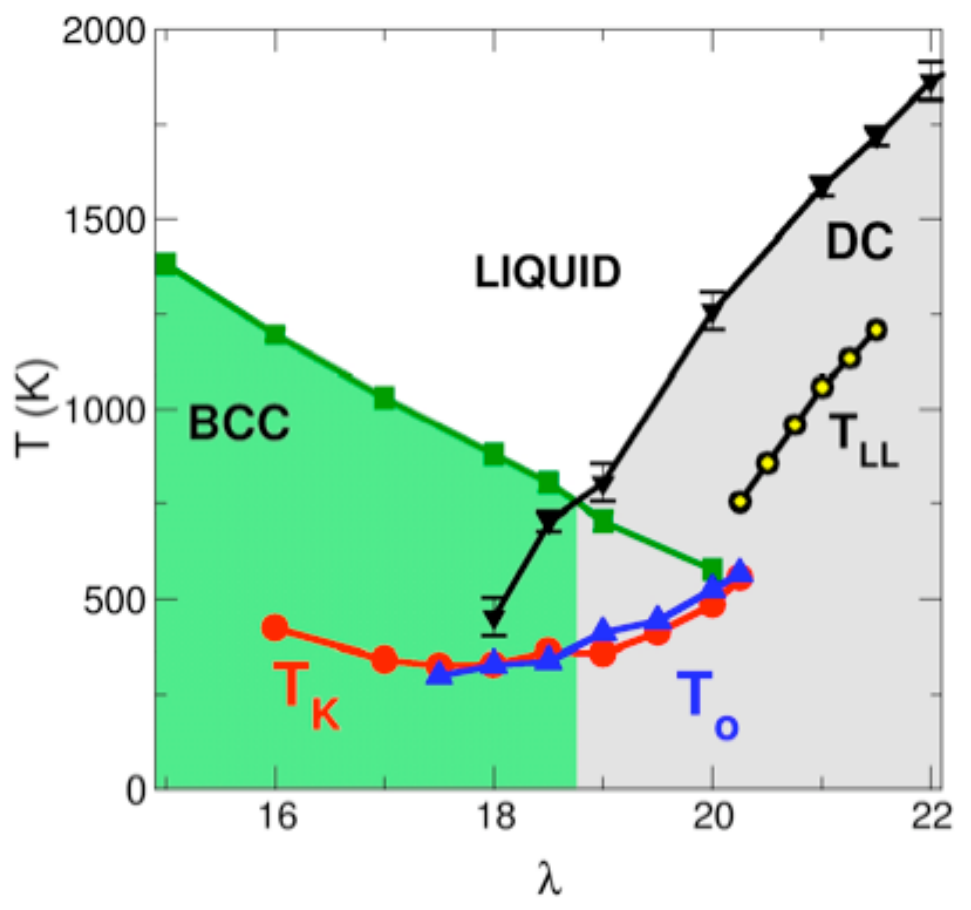

**Figure 1.** The temperature potential phase diagram for the λ-modified Stillinger-Weber silicon system[13]. The λ value for silicon is 21. The enthalpy of (metastable) melting of the DC crystal with l = 18 is close to zero, and for smaller lambda values becomes positive. Crystals with λ values in the glassforming range may be obtained by changing the potential while in the crystalline configuration. From ref. 13, by permission).

The liquids that formed near the minimum melting point had a diffusivity equal to that of the experimental system $Ni_3P$[20] (viz., 0.95 x $10^{-9}$ $m^2s^{-1}$) and proved incapable of crystallizing on the longest computational time scales at any temperature below this pseudo triple point. Indeed, over a range of potentials around that giving the minimum melting point, no crystallization was observed. This seemed to support the idea that a certain low diffusivity at $T_m$ would be enough to ensure slow crystallization kinetics and glass formation. The low melting point itself could be associated with the reduction in the enthalpy difference between liquid and crystal states. For the λ = 19 parameter, the lattice energies of the competing crystals were essentially identical, though the enthalpies and entropies of the two crystals differed at the melting point due to different anharmonicities. An important feature of the pseudo-triple point potential is that at this potential the excess free energy of the liquid, hence the thermodynamic drive to crystallize, increases most slowly with increasing undercooling.

On the other hand, earlier studies of the original SW silicon potential[21, 22] show that slow diffusion alone does not ensure slow crystallization. When SW silicon, $\lambda = 21$, is supercooled to 70% of its melting point, it remains liquid over the longest simulation times. But when the temperature is lowered 1% more, and the liquid-liquid phase change occurs at constant free energy, the new liquid phase proceeds rapidly to crystallize despite being three orders of magnitude *less* diffusive. Since the free energy drive to crystallize has not changed, what must be responsible for the rapid nucleation is a dramatic drop in interfacial (liquid-to-DC crystal) tension[23]. Such a drop occurs because, in the LL transition, the liquid phase topology has become much closer to that of the crystal. However much difference remains, as can be seen from the fact that enthalpy of crystallization remains much larger than the enthalpy of the LL transition. That the $\lambda = 19$ liquid phase should remain uncrystallizing at the same liquid diffusivity $0.95 \times 10^{-12}$ $m^2s^{-1}$ as the rapidly crystallizing silicon low temperature liquid phase, can be understood from the great difference between the liquid structure and the structures of the two competing crystalline phases. This can be seen in the radial distribution functions[13](supplementatry information). When the structures are different the interfacial tensions remain high, hence none of the conditions favorable for crossing the nucleation barriers are being met).

The direct follow-up of these studies has been the successful laboratory vitrification of an analog of the mSW19 liquid[24] . The analogy is in phase diagram, not in the structure. Being a laboratory study, it is constrained to the periodic table and the element chosen for study was germanium. To obtain the melting point lowering, pressure was substituted for potential tuning. Diamond anvil high pressure cells were used to obtain the pressure while permitting melting by focused $CO_2$ laser beam short term pulses, just sufficient to fully melt the sample[24]. The vitrification could not be observed by in situ X-ray study due to the small sample size. However if a crystalline high pressure metal phase had formed during cooling under pressure, Bragg peaks of that phase could have been observed but were not seen.

The electron micrographs and diffraction patterns used to confirm the vitrification of Ge, are shown in Figure 2. The most interesting feature of Figure 2 is actually the globules, which were not seen in successful vitrifications at higher pressures. These have all the appearance of globules seen in compositionally driven liquid-liquid phase-separated metallic glasses reported by Hono and coworkers[25, 26] – though in our case of course there is only one type of atom. The globules were seen only in samples cooled in the pressure range 7.6-7.9 GPa near the triple point.

The interpretation of these findings is illustrated in Fig. 3. The point to emphasize here is that globules are trapped in the quenching at 7.5-8 GPa because at this pressure the glass temperature and LL phase transition temperature overlap, so that globules of the low temperature liquid phase are trapped as they form. Most of the globules are found to be fully amorphous but some have crystals of the diamond cubic phase that have grown within them. These are the only locations in which crystals are found. This is fully consistent with the observations made earlier on the MD simulations of silicon[21, 22] concerning crystallization where crystals were only found to form in the low temperature phase.

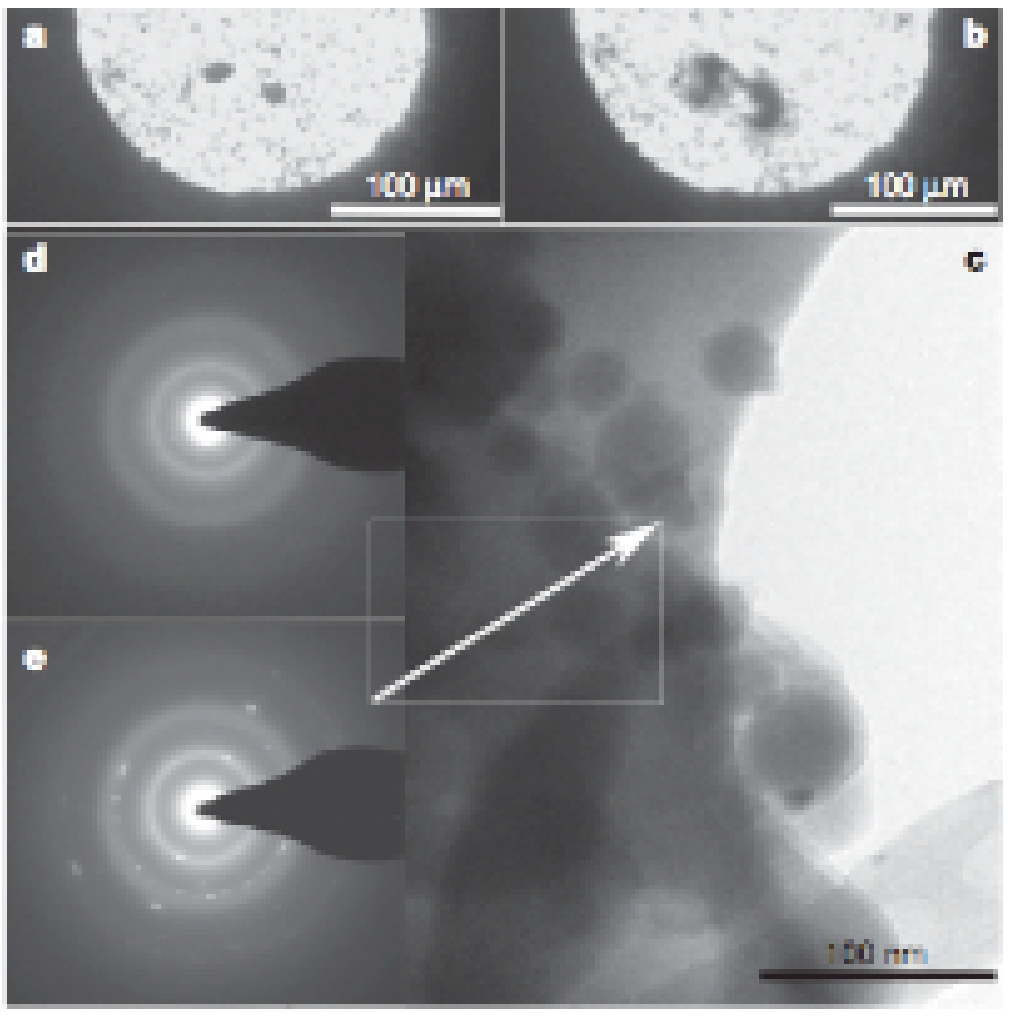

**Figure 2.** (a) and (b): light microscope images of diamond cell aperture showing Ge sample (dark) both before (a) and after (b) the pulsed melting. Part (c) shows a TEM of part of the after-quench sample seen in (b). The image reveals many globules which it is believed formed by liquid-liquid transformation during the cooling, just as the system was becoming trapped in the glassy state. Part (d) shows purely amorphous halos that are characteristic of most parts of the sample, but part (e) shows presence of some crystalline material. High resolution TEMs (see ref. [24]) show that the crystal is growing out of the globule at the pointer tip. (Adapted from ref. [24] by permission)

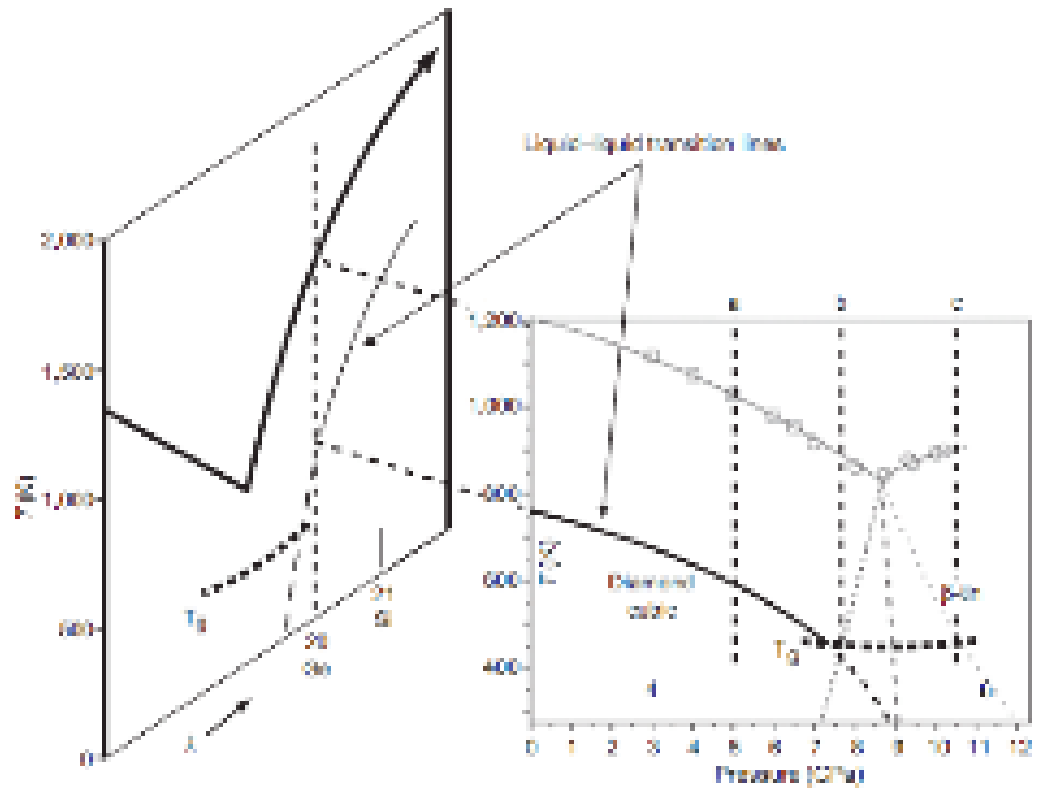

**Figure 3.** Phase diagram of Ge under pressure, shown in projection from the T vs. potential (l) diagram of ref. [13]. Not particularly the position of the liquid-liquid transition line in the temperature vs pressure diagram, and the manner in which droplets of the LDL phase formed during cooling t 7.5 GPa (line b) must get trapped in the glassy state as they form (since the low entropy phase is less diffusive, by the Adam-Gibbs equation). (From ref.[24], by permission of McMillan Pub.)

The low temperature liquid phase appears to be an "Ostwald step" along the decreasing free energy path to the stable crystal phase. Nucleation to the crystal proceeds more quickly along this path, despite a lower diffusivity, because of the lower interfacial free energy barrier. In how many different types of systems this phenomenon will be encountered, remains to be seen. So far, besides the case just described of a metallic liquid (en route to a covalent crystal), examples have been found in the field of ionic liquids (yttrium oxide-aluminum oxide - the first case reported[27]), and molecular liquids (tri-phenyl phosphite[28]). The case of $GeSe_2$ studied by Crichton et al[29] might provide an example from the chalcogenide glass field.

**The Jagla model, its derivatives, and its laboratory manifestation.**

Although it is the modified SW potential that has so far received the most attention with respect to crystallization studies, there have been some interesting reports on melting and supercooling relations in a system of even simpler potential. This is the two-scale Jagla model[30] which features a spherically symmetric potential with discontinuities. The model has a hard core, and a ramped repulsive wall, and the ratio of ramp length to hard core can be used to tune the model properties which include certain water-like anomalies[31]. When the model is given an attractive well, as in the studies of Stanley and coworkers[32], and of Gibson and Wilding[33], the liquid phase develops a second critical point. The value of the critical temperature relative to the melting point can be tuned using the ratio of length scales, (at constant second virial coefficient, in Gibson and Wilding's study) and the slope of the liquid-liquid coexistence line can be changed from positive, like OTP, to negative (as proposed for water) by change of this ratio. Also changing with this ratio is the relation between critical point and melting point. Values giving water-like slopes and submerged critical points give rapid crystallization while those with positive slopes, and liquid-liquid critical points in the stable state, yield good glassforming ability for *each* of the two liquid phases[34] (though the resistance to crystallization varies with the pressure acting on the system).

Continuous versions of this potential have now been developed[35] , and these are found to have essentially the same properties as their less physical, discontinuous potential, counterparts.

An exciting aspect of the Jagla model is that it has an experimental counterpart. This is a system that can itself be tuned, using the same pressure variable that was applied in the study of laboratory germanium discussed above. We refer to the element cerium, Ce. Ce has an interesting electronic structure, which is responsible for its widespread use in redox systems ($Ce^{3+}/^{4+}$), but it is the elemental state in which we are interested here. Due to the energetic similarity of its electronic states near the Fermi level, and to the different spatial requirements of these states, the atomic volume of condensed phase Ce can change spontaneously when (at a given temperature) $P\Delta V$ exceeds the (zero pressure) orbital energy difference. It is as if pressure squeezes one of the valence electrons out of the electron sea "conduction band" back to the ion core (valence band). The atom-atom interaction potential describing such a system should have a spherically symmetric soft repulsive wall, like the Jagla model.

The experimental exploitation of this similarity for liquid state studies has yet to be undertaken, and may not be simple for the following reason. A related transition, FCC <-> FCC, is known to occur in the crystalline state of Ce, indeed this was the first recorded example of an isosymmetric phase transition and the first case for which a theory was developed to explain the observations[19]. Figure 4 shows how this transition transition may be observed over a range of pressures with a predictable Clapeyron slope[19]. It terminates at a critical point some 100's of K below the melting point – which immediately raises problems for its observation in the liquid. The physical state of the substance should not greatly affect the temperature at which the transition occurs so its observation in the liquid would require supercooling. Pure liquid metals can usually be deeply undercooled[36] though close to a critical point this would probably change and supercooling propensity would likely show a minimum near the critical pressure. (Thus we see how an interaction potential could be tuned to *minimize* glassforming propensity, as well as the converse).

In view of the above considerations, modification of the melting relations by appropriate second component additions may be the key to revealing interesting behavior, with the emergence of a second critical point into the stable liquid state, a remote possibility. Indeed there is a recent report of a smeared polyamorphic transition in a glassy Ce-Al alloy which is 55% Ce[37], and $Ce_{70}Al_{10}Ni_{10}Cu_{10}$ is a BMG with unusual properties[38]. Studies of the effects of second components on the Jagla liquid have recently been made, with interesting predictions[39] which, we now see, may apply to cerium binary liquid alloys under pressure.

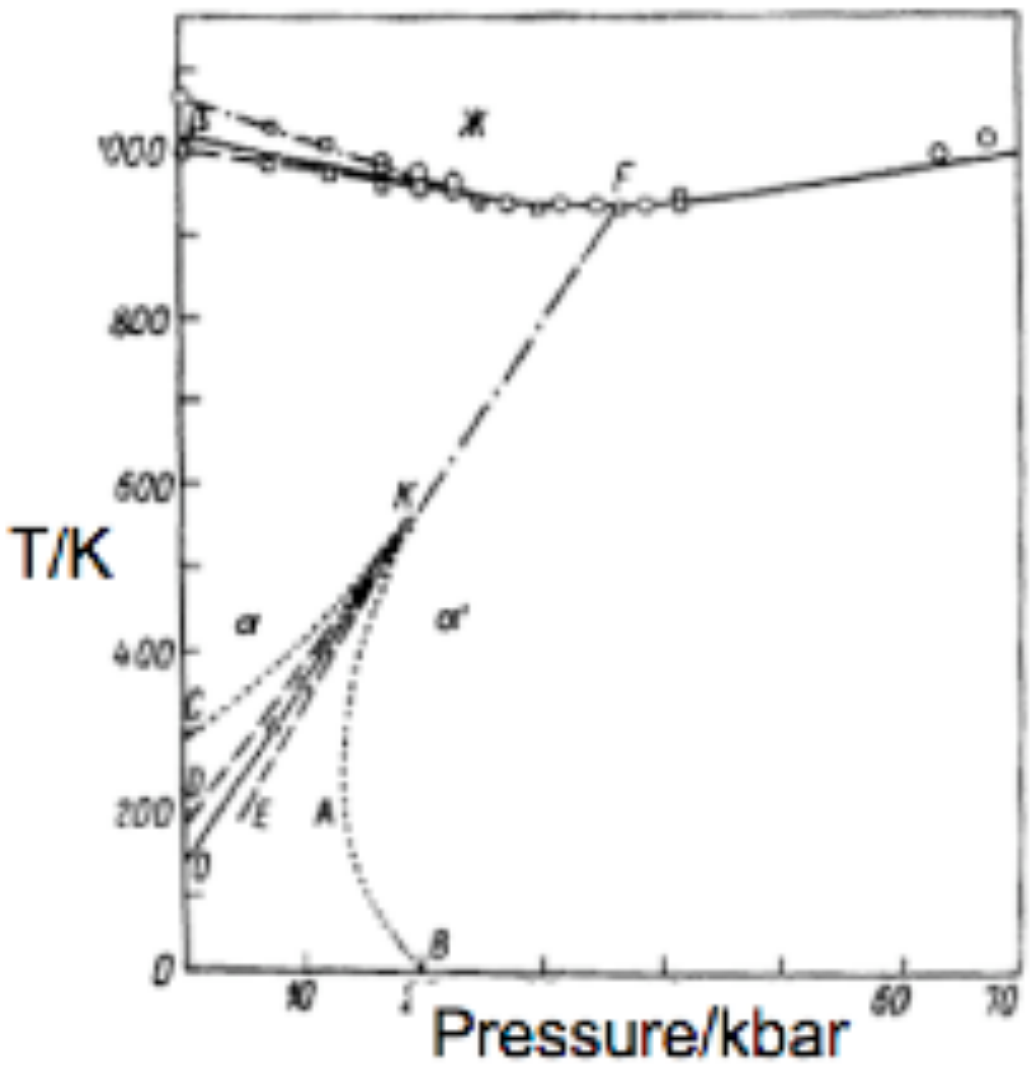

**Figure 4.** Phase diagram for the element cerium showing how pressure affects the α–α' (FCC(1)-FCC(2) isosymmetric phase *transition*. Point K is a critical point. Cooling from above this point at constant pressure should produce a lambda-like anomaly like that seen in the final figure of this paper, and like that seen in the liquid state simulations of the two-scale Jagla model. Whether or not there would also be an ergodicity-breaking transition at lower temperatures (like that in the Jagla model and that in the final figure of this paper) is an interesting question in view of the electronic degrees of freedom involved in the transition. (after Aptekar and Ponyatovsky ref. 19.)

## 3. PROPERTIES OF THE VISCOUS LIQUID PHASE.

Once the supercooled liquid state has been opened for study, a number of provocative behavioral features are revealed[40], and these will be briefly reviewed before asking some broader questions.

The first feature is the appearance of major deviations from the Arrhenius rate law, for relaxation times τ and transport properties like diffusivity, that is so familiar in (i) chemical rate processes and (ii) the physical responses of most condensed phases to perturbations from the equilibrium state. The manner in which a variety of equations, (either empirical, or based on one or other physical model) can account for the deviations from Arrhenius behavior, has been described in some detail in the literature[41-43]. The most commonly applied is the 3-parameter Vogel-Fulcher-Tammann equation,

$$\tau = \tau_0 \exp(T_0/F[T-T_0]) \qquad (1)$$

where $\tau_0$, F, and $T_0$ are constants, but there are other variants on the Arrhenius equation with comparable numbers of free parameters, that perform as well, e.g., the

Avramov equation[44],

$$\tau = \tau_0 \exp B(T_g/T)^n \qquad (2)$$

where $\tau_0$, B, and n, are constants. The parameters F (Eq. (1), and n (Eq. (2) determine the departures from Arrhenius behavior that are seen in the composite Arrhenius plot, Fig. 4(a)[45] and thus determine the sensitivity to temperature change, now generally known as the liquid "fragility".

In addition, there are models in which success is achieved with fewer parameters by invoking another physical property, usually a thermodynamic quantity, such configurational entropy, $S_c$, in the Adam-Gibbs equation[46],

$$\tau = \tau_0 \exp(C/TS_c) \qquad (3)$$

or the high frequency shear modulus $G_\infty$ in the shoving model[47]

$$\tau = \tau_0 \exp(AG_\infty/T) \qquad (4)$$

where $\tau_0$ is a constant and the temperature-dependence of these thermodynamic quantities explain the deviation from Arrhenius behavior.

More recently, it has been suggested [48] that the thermodynamics enters the relaxation time expression through the fluctuations in energy between potential (configurational) and kinetic (vibrational) manifolds which build up the configurationally "hot" domains via which relaxation occurs. This leads to an expression for the relaxation time of Eq. (1) form, but without divergence. It is

$$\tau = \tau_0 \exp( DT'T/[T^2 - z\lambda T']) \qquad (5)$$

where D and T' can be considered as fitting parameters.

The thermodynamic connection can be highlighted by showing the temperature dependence of the thermodynamic quantity in a form comparable to Fig. 5(a) This is shown for the available quantity, the excess entropy of the liquid over crystal[45], in Fig. 5(b) for each of the substances of Fig. 5(a) using the same scaled inverse temperature representation of their variations. Fig. 5(b) graphically makes the point that *the origin of whatever is found interesting in the behavior of the viscosities of Fig. 5(a) is likely to be found in the thermodynamics of these systems*. The challenge is then to explain the thermodynamics, which will be taken up below. First we summarize some of the more subtle aspects of the relaxation phenomenology. These lie in the details of the relaxation process, in particular in the deviations from its usual exponential character, that are found whenever the relaxation function is determined. Such studies have been the focus of a great deal of research activity in recent years, in particular the connection to dynamic heterogeneity identified initially in simulation studies by Hurley and Harrowell[49]. Progress has been reviewed in detail by Ediger[50] and Richert[51] and we summarize only enough to take the next step in the quest for thermodynamic understanding.

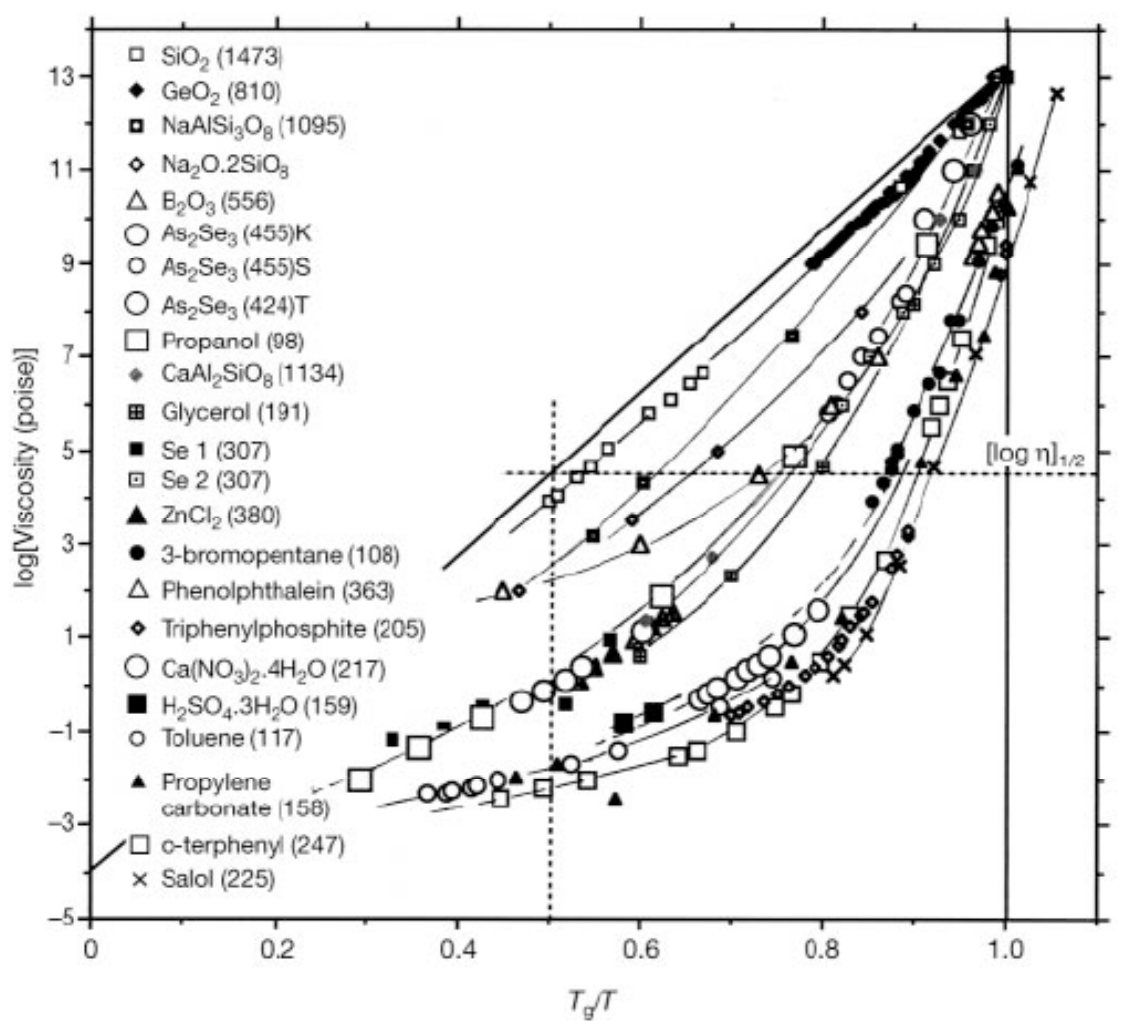

**Figure 5(a).** $T_g$-scaled Arrhenius plot of viscosities using “onset” $T_g$ values determined by differential scanning calorimetry at 0.33 K/s.

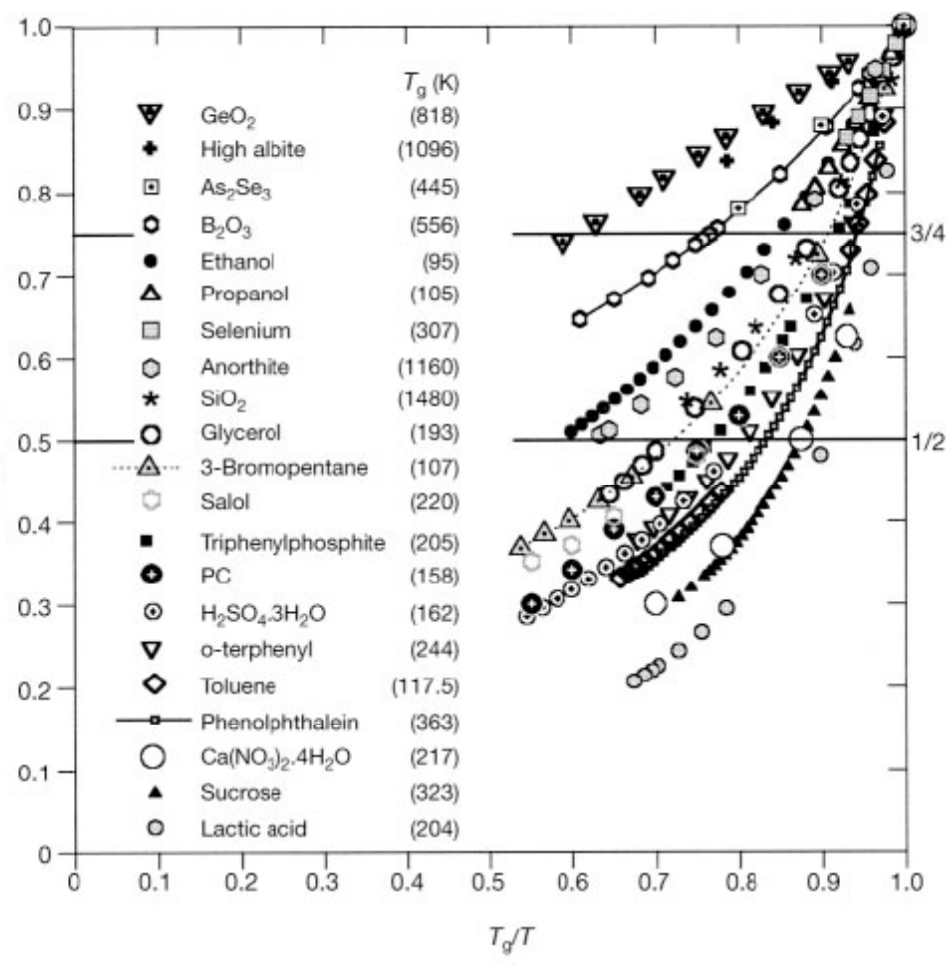

**Figure 5(b).** $T_g$-scaled inverse temperature display of the increase of excess entropy of liquid over crystal as temperature rises above the onset glass temperature $T_g$

**Non-exponential relaxation and dynamic heterogeneity.**

At temperatures characteristic of simple liquids, or of glassformers near their boiling points, the return to equilibrium after some perturbation (or, equivalently, the decay of spontaneous fluctuations about the equilibrium state) is found to be exponential in character. In this temperature regime the systems’ relaxation times also obey the Arrhenius law. However, it appears that at the same point in temperature at which the non-Arrhenius regime is entered from above, glassforming systems develop a heterogeneities in their dynamics. The heterogeneities are such that , for a limited period of time, one set of particles will lock together while an adjacent set will become loose and (according to both MD simulations[52] and colloidal particle studies[53]) will support string-like motions of the particles. Within each nanoregion, relaxation appears to be exponential[51] , and it is therefore the distribution of nanoregions that determines the deviation from exponentiality seen in the macroscopic relaxation function[51].

In certain unusual systems, particularly normal alcohols studied by dielectric relaxation, the relaxation function can be found to remain exponential while the relaxation time is non-Arrhenius[54]. In these cases, which include water [55] it is always found that the dielectric process is slower than the structural relaxation, by factors of 10-2000[56] . According to recent work[57], it is the latter, alone, that carries the calorimetric strength i.e. $\tau_D > \tau_s \sim t_H$ (see ref. [57]).

The exact nature of the heterogeneities and the question of how directly, if at all, they relate to the structure has been a matter of controversy. Investigation of the latter question by Harrowell and coworkers[58, 59] has identified a “propensity” for fast relaxation that is embedded in the structural organization of the particles. This has suggested that the nanoheterogeneity might be described as a dynamic *nanogranularity*. Most recently, in a development of the dielectric hole-burning studies instigated by Chamberlin in a series of proposals[60] and then successfully demonstrated by Schiener et al.[61], Richert and coworkers[62, 63] have been able to provide new insights into the thermodynamics of glassformers. As the latter authors emphasize[63], these nonlinear experiments directly excite the configurational manifold, and the experimentalist then observes the (relatively) slow leakage of energy from the excited configurational states back into the phonon bath (the opposite direction of energy flow from that in "normal" experiments). By modeling this effect, they have been able to show that the thermal and dielectric time constants are locally correlated and, especially[63], that the measured excess heat capacity of liquids is only partly configurational in character – as had long ago been inferred by Goldstein[64]. Furthermore, they find that the configurational fraction tends to be smaller in fragile than in non-fragile liquids, as conjectured more recently[45, 65].

These interesting findings are in need of confirmation by study of systems that are less complicated than the great majority of glassforming liquids tend to be. Fortunately there are possibilities, involving other systems, that have yet to be exploited and these will be discussed in the next main section. In order for the usefulness of this excursion into other systems to be properly appreciated, we first need to take a broader look at the thermodynamics of liquid glassformers (remembering from Fig. 5 that the broad pattern of strong/fragile glassformer behavior is imprinted in the thermodynamic properties of glassformers as clearly as it is in their dynamics).

**Excess heat capacity behavior across the broad spectrum of glassformers**.

Glassformers of common experience share the phenomenon of abrupt heat capacity drop when ergodicity is broken during cooling. It is the usual way of defining a glass transition. The shape of the heat capacity function that is interrupted at the glass transition is, however, subject to broad differences as illustrated in Fig. 2 of ref.[40]. In metallic glassformers, the increase in heat capacity is particularly sharp: in the more fragile cases the heat capacity increases by a factor of 2 or more from the classical Dulong and Petit base, and then drops very quickly on further rise of temperature. On the other hand, in the classical network glasses the trend is the opposite, the heat capacity tending to increase with increasing temperature above the glass temperature.

In $SiO_2$ that is free of water, the heat capacity jump at $T_g$ (1200ºC) is very small. Its behavior above $T_g$ is difficult to guage because of the high temperatures involved, but the case of the weak-field analog of $SiO_2$, $BeF_2$, it is easier to evaluate because $T_g$ is only 319ºC. In this case the heat capacity jump at $T_g$ was too small to record by the drop calorimetry method used in its most extensive study [66, 67] but the continuous increase above this temperature was unambiguous. Data were reported from 350-1000K by which temperature $C_p$(ex) had increased from negligible to 25% of the classical vibrational value of 3R/g-ion. When MD results are included [68] it is found that there is actually a peak in heat capacity (at ~1.5R) a little above the high temperature limit of the experimental study. A more complete MD study has since been reported by Scheidler et al. [69] for the case of $SiO_2$ in the BKS potential. In this study, the real and imaginary parts of the specific heat were calculated from the temperature fluctuations at equilibrium. The static values are reproduced in Figure 6, where an increase from very small (extrapolated) values at the experimental $T_g$, towards a maximum in the vicinity of 5000K, is seen. When scaled by the $BeF_2$:$SiO_2$ $T_g$ ratio, this maximum would fall on that observed for $BeF_2$, seen in the insert. Thus a certain pattern for network liquids begins to emerge.

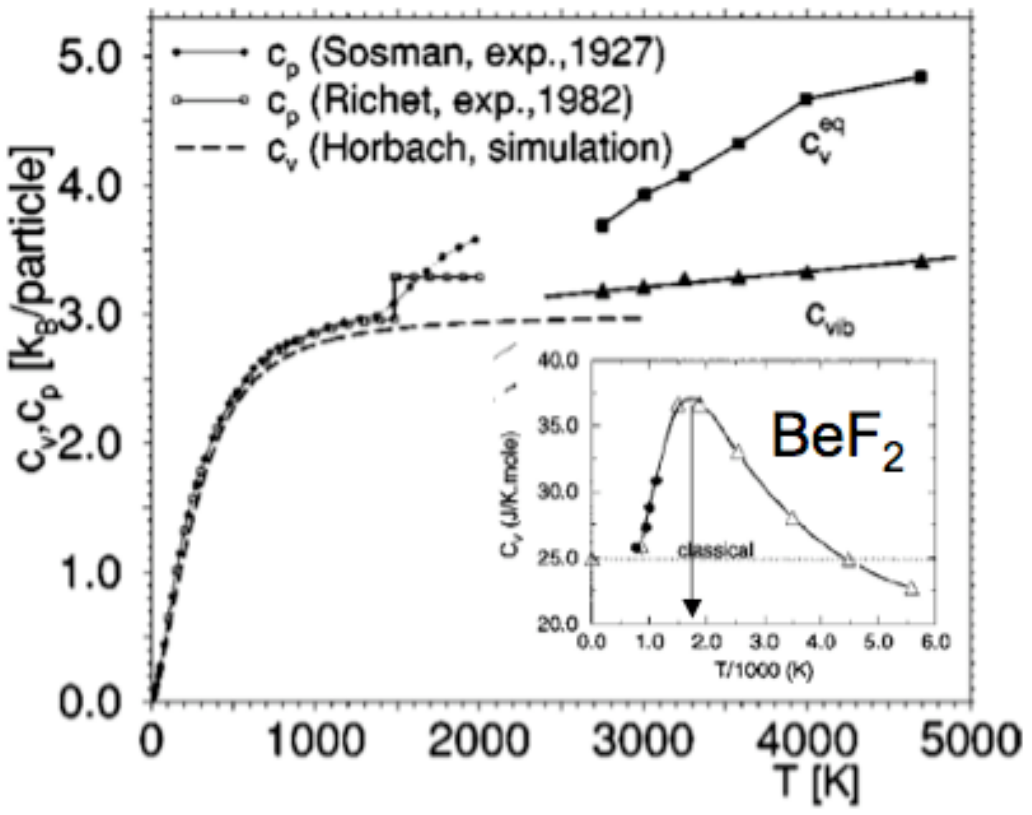

**Figure 6.** The heat capacity of liquid and glassy silica calculated in the MD simulations of Scheidler et al[69], compared with data from experimental and earlier simulation studies on $SiO_2$. Inset: The findings for $BeF_2$, both experimental (•) and MD simulation (o). Scaling by the $T_g$'s, (570K and 1273K), a maximum at 4700K for $SiO_2$ would be expected from the $BeF_2$ data if the two are equivalent in behavior.

In Figure 7, these findings are put together with those for molecular liquids, and those for water (discussed elsewhere[70] ), in an attempt to construct a "big picture" for glassformers. The glass transition for water, vitrified by three different procedures, is in each case so weak that its existence has been the source of controversy for decades[71] . A recent rationalization of this weakness has been that the heat capacity for this hydrogen bonded tetrahedral network liquid is distinct from that of normal molecular liquid glassformers, and belongs instead to a transition of the lambda type, which we discuss further below, and that the ergodicity-breaking occurs in the tail of the transition where there is little heat capacity left to lose[70]. The heat capacity spike of the lambda transition would in that case occur below the melting point and could well have first order transition character, but this cannot be seen because of the prior occurrence of crystallization. Crystallization may be promoted by the large entropy fluctuations associated with the heat capacity spike – or may be even more directly promoted by a silicon-like liquid-liquid transition to a rapidly crystallizing LDL form (see first section of this paper).

Continuing the progression from smeared peak located above the melting point in ionic network glasses to sharp peak below the melting point in the case of hydrogen-bonded water, we find the cases of glassforming metals and fragile liquids in which there are no peaks at all but only increasing heat capacities until ergodicity is broken. According to the Gaussian excitations model[48], this is due to the peak now falling below the glass temperature or, more probably, being replaced by a weak first order transition to the ground state. This model predicts the transition temperature to lie below a critical temperature $T_c = \lambda/2$, where $\lambda$ is the disorder stabilization energy for excitations, at the temperature

$$T_{LL} = (\varepsilon_0-\lambda)/s_0$$

where $\varepsilon_0$ is the basic (pre-stabilization) excitation energy and $s_0$ the excitation entropy, which must be greater than $2k_B$ per bead (rearrangeable unit) for the transition to exist. It is a puzzle that there are so few cases of systems with liquid-liquid transitions above $T_g$. Kurita and Tanaka[72] have shown that Kivelson's "glacial phase" of tri-phenyl phosphite, TPP is such a case and anticipate[73], as do we[48], that there should be many others but have so far only identified t-butanol as additional. The LL transition for TPP is indeed associated with rapid crystallization, as is the case with other polyamorphic transitions[22, 24].

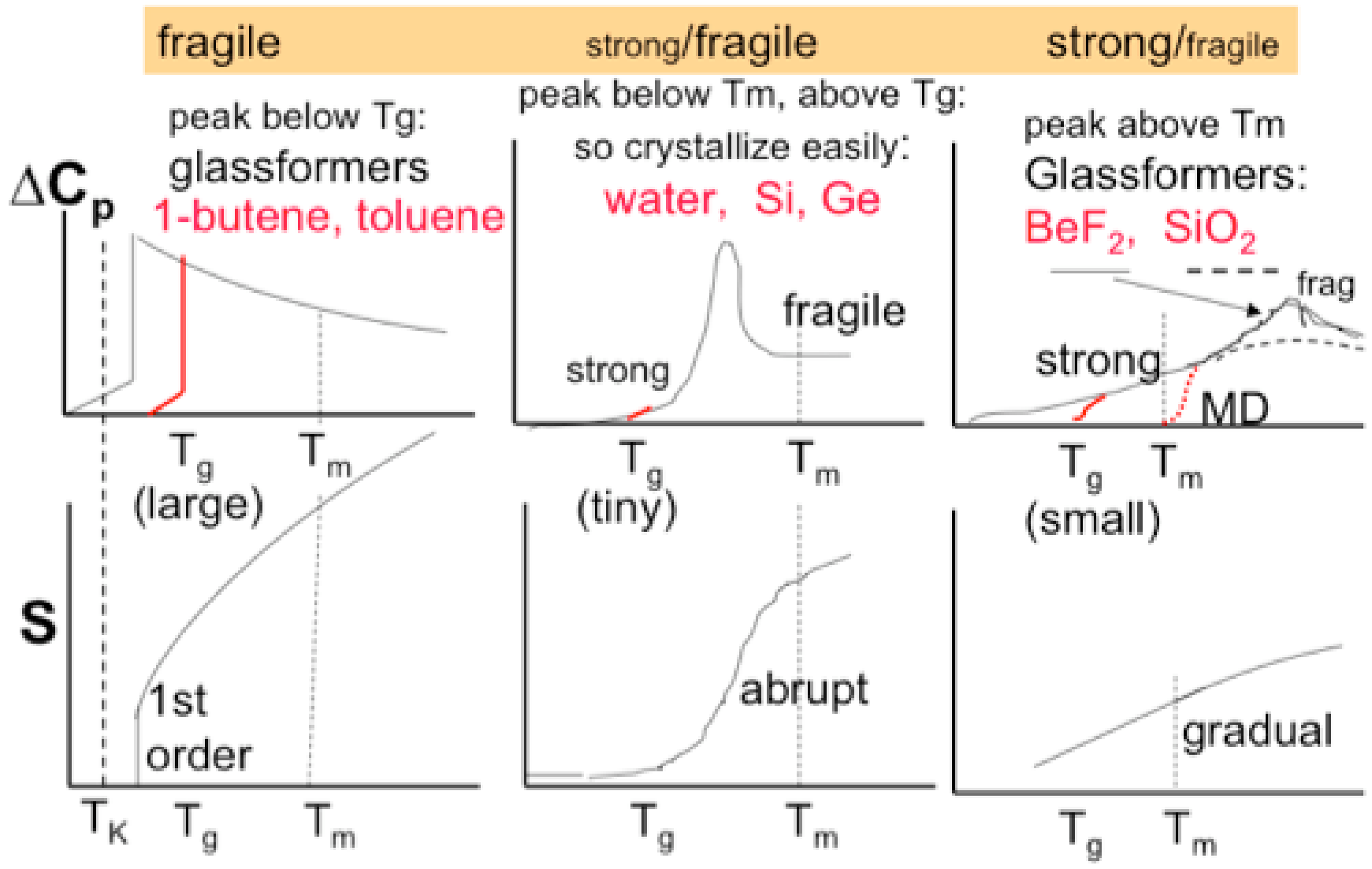

**Figure 7.** Changeover in the heat capacity form of the glass transition, on passing from "strong" inorganic network glasses to "fragile" molecular glasses. This is to be interpreted in terms of the increasing Gaussian width in the distribution of excitation energies, and consequent increasing disorder stabilization of the excitations, leading to an increasingly cooperative excitation process from networks to molecular liquids.

## 3. THE RELATION OF GLASS TRANSITIONS IN LABORATORY LIQUIDS TO GLASS TRANSITIONS IN LABORATORY CRYSTALS.

The essence of a glass transition is the change of properties that occurs when the time scale for a temperature-dependent (or pressure-dependent) degree of freedom of the system exceeds the time scale of the experiment being used to study the system. Thus there are glass transitions in magnetic systems (spin glasses) and superconducting solids (vortex glasses) and there are different sorts of "structural" glasses[74]. The structural glasses formed from liquids have provided the material for this article up to this point, but we will profit from a discussion of the structural glasses that occur within crystals, particularly some that until this time have been more or less ignored by the field.

The existence of orientational glasses, called "glassy crystals" by Seki and co-workers who first investigated them[75], is generally recognized, and their high degree of similarity to the liquid-formed glasses (non-Arrhenius temperature dependence, non-exponential relaxation functions, Kauzmann temperatures, etc) has now been described by many workers[76-79]. It is observed that, relative to glassforming liquids, the glassy crystals exhibit much "stronger" behavior. Arrhenius behavior is relatively common, whereas it is rare amongst liquids.

The glass transitions in the more fragile glassy crystals provide many of the same challenges offered by the molecular liquid glassformers on which most of the work in the field has been carried out, and in some cases – for instance, ethanol – there is little phenomenological difference between them at all. As an example, the data on ethanol in Kauzmann's famous plot[80], were later found to be data for the ethanol glassy crystal[77]. On the other hand, there are glass transitions (in the sense of ergodicity-breaking processes) that occur in disordering crystals that seem at first sight quite different from "glassy crystals", and so bear very little resemblance to the molecular glass formers on which most of the work in the field has been done. We discuss these in the next section.

**Glass transitions as the kinetic cutoff on lambda (order-disorder) transitions.**

Although the terms "lambda transition" and "order-disorder transition" have now acquired precise statistical mechanical meanings within certain of the "universality classes" of critical phenomena[81], we will use the terms here in their original broader sense, i.e., to describe transitions that exhibit an accelerating heat capacity that peaks sharply (without any first order , hysteritic, character) before decreasing sharply to a much lower, usually phonon-determined base-line. The ergodicity-breaking phenomena in which we are interested here always occur at temperature that are far below the peak values (logarithmic singularities) at which fine universality class distinctions, based on the values of "critical exponents", are made. Thus the precise universality class of the individual transition is not of much import to the present paper. Indeed, in some cases of great interest to us, the system will never reach a higher order transition point but will instead encounter a first order transition at which the disordering process is abruptly concluded. This will not decrease the interest content of the ergodicity-restoring transition (glass transition) that we can find occurring during reheating before the first order transition temperature is reached.

The disordering process in these transitions usually occurs over a wide temperature range, over most of which the system remains ergodic, such that very little residual entropy is found frozen in at OK. Thus the energy landscapes representative of systems exhibiting lambda transitions must be very different from those characterizing most structural glassformers. To understand how so little entropy is frozen in at the glass transition temperature, it must be supposed that the energy barriers separating states on the landscape are very small and that their arrangement conforms to the "palm tree" disconnectivity graph described by Wales[82, 83]. Then the system can remain ergodic over almost the whole of the excitation profile. This provides a contrast with the normal molecular glass-forming substance whose configurational heat capacities are peaked in the other direction. Indeed the peak value of the configurational heat capacity of the normal glassformer is determined by the breaking of ergodicity, thereby ordaining that the full form of the excess heat capacity is never revealed.

Weak ergodicity-breaking is occasionally reported in the low temperature tails of lambda transitions, though not much interest in such "glass transitions" has been evoked. One of the more notable cases is that of the fullerene, $C_{60}$, that was thoroughly studied by Matsuo et al[84]. The disordering relaxation times in $C_{60}$ have also been determined using a variety of techniques, most extensively by dielectric relaxation[85] and were shown strictly to obey an Arrhenius law. This case was recently used by the author[70] as an example to help rationalize the abnormal and controversial glass and supercooled liquid state behavior of the important substance, water.

Of much greater interest to us in this article, however, are the cases of certain metallic superlattices in which the kinetics of disordering are evidently much slower than usual, so that ergodicity is broken long before the ordering process is completed. An example from the literature of the first half of last century is provided in Figure 8. Here the body-centered cubic lattice of iron is occupied by both iron and cobalt that, at lower temperatures, are ordered into two interpenetrating simple cubic lattices. As temperature increases, the elements begin to exchange places until at 700ºC (well before the BCC-to-FCC transition so familiar in pure Fe) the disordering is completed. The heat capacity behavior of this simple system was reported in 1943 by Kaya and Sato[86] and their findings are reproduced in Figure 8. Why these data have failed to attract much attention is something of a mystery.

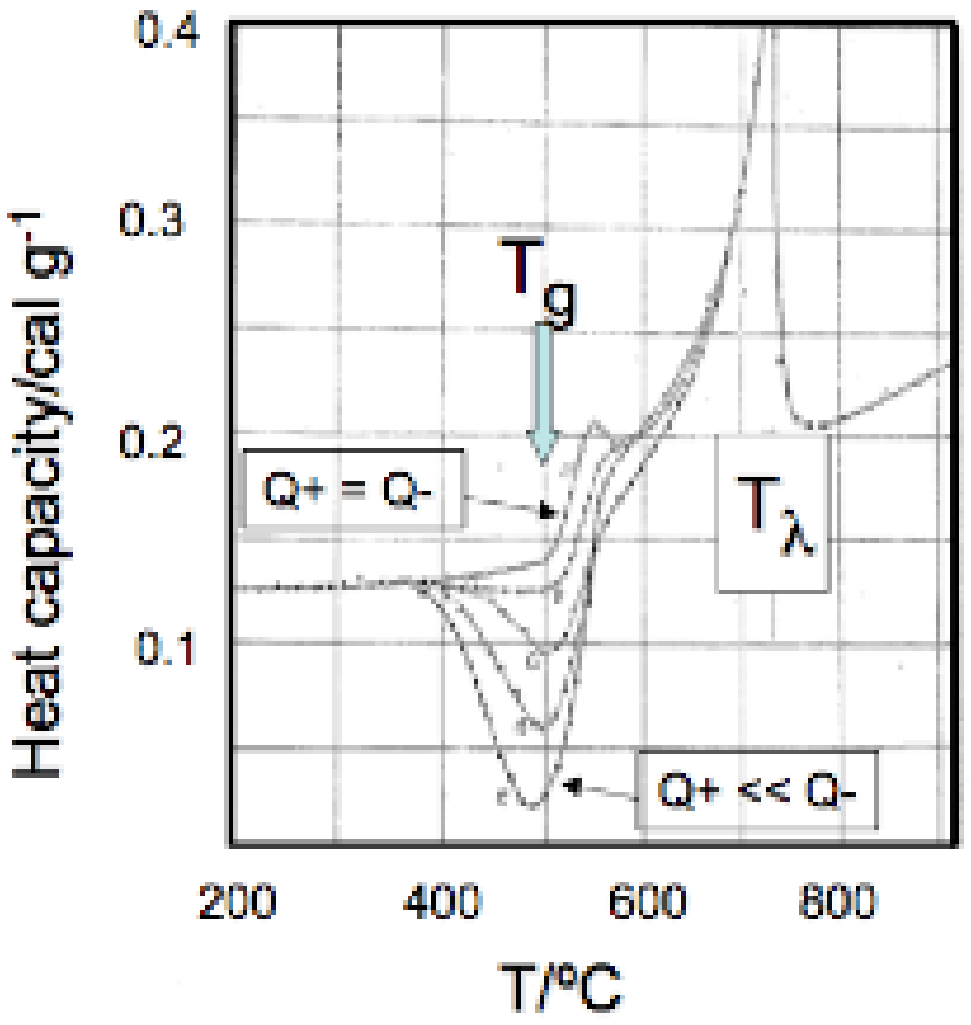

**Figure 8.** The heat capacity of Fe-Co alloy of atom ratio 1:1 as a function of cooling rate from various temperatures above the glass transformation (relaxation time-dominated) range. The exothermic components observed for faster cooling rates can be area-matched to the equilibrium heat capacity to estimate how the heat capacity measured during cooling at the different rates would appear if it could be measured directly. The case Q+ = Q- is the "standard" glass transition for this system. Note that an even simpler system showing a transition of this type exists, and has already been presented in this paper. It is the element cerium at the pressure of point K on the phase diagram, Figure 4.

The behavior of the heat capacity scans obtained on samples that were cooled much more rapidly than the upscan rate is close in character to that found in the scanning of quenched and hyperquenched molecular glass formers[87], and clearly is to be associated with the higher levels on the energy landscape at which the system is trapped during the faster cooling processes (Q->>Q+). An important difference between these studies and those on the liquid glassformers is that the equilibrium form of the heat capacity for the alloys is not an unknown. The equilibrium transition is of the lambda form, and there is no suggestion of anything else, i.e. there is no Kauzmann paradox or other related mystery to ponder.

The behavior of the heat capacity scans obtained for cooling rates (Q-) greater than the heating rates (Q+) of the heat capacity scans, is close in character to the behavior found in the scanning of quenched, and hyperquenched, molecular glassformers[87, 88]. It is clearly to be associated with the escape from the shallower traps at higher levels on the energy landscape in which the system became arrested during the faster cooling rates.

What is desirable to know is the relation of the disordering kinetics in these alloy systems to those characteristic of the ordinary (liquid) glassformers. Are the dynamics heterogeneous as in the common glassformers and in the glassy crystals and do they (inverse to normal glassformers) become less heterogeneous as they approach their weak glass transitions in the alloy cases? It is possible, for instance, that the heterogeneity reflects the magnitude of the fluctuations rather than the magnitude of the disorder. We note that strong liquids like $SiO_2$ exhibit stretched relaxation kinetics despite their low fragilities. Likewise, the chalcogenide glasses, though exhibiting greater stretching when fragile, level off at $\beta = 0.65$ on passing to less fragile compositions[89].

The final observation to be made is that, from the point of view of heat capacity forms, the network liquids would seem to have more in common with the crystalline state lambda transitions than they have with molecular glassformers. Are the strong liquids of glass science no more than early-stage disorder-perturbed lambda transitions? The challenge here, then, will be to understand how, simply as a consequence of chemical modification (by adding network modifiers that disrupt the bonding organization), the network liquids can become phenomenologically indistinguishable from the molecular liquids. Also we must ask, are the stronger plastic crystals merely order-disorder transitioning systems that melt before they reach their lambda peaks? The case of $C_{60}$ is, after all, unusual insofar as $C_{60}$ (due to its hard sphere-like character) cannot melt, but rather can only sublime at higher temperatures[90].The sublimation temperature is higher than the critical point for the (always metastable) liquid. Clearly, there is need for the study of chemical disordering in plastic crystal lattices which should be possible in most cases because the disordered crystals are generally more tolerant of second components than are their more ordered cousins.

**Concluding remarks.**

To limit the complexity of discussion in this article, we have supposed that only one type of disorder dominates the thermodynamics and dynamics of each of the system types we have considered. In reality, systems tend to be more complicated, and it is possible to have a disorder in one degree of freedom, e.g. orientation, that has different kinetics from those of other degrees of freedom, e. g. those that involve a disorder in chemical environments. In complex aluminosilicate liquids, for instance, it is certain that the equilibration of the Al/Si distribution proceeds on time scales quite different from, and slower than, the shear relaxation time for the liquid. Thus there should be more than one ergodicity-breaking process occurring in complex liquids as the cooling from high temperature regimes (where all relaxation times are very short) proceeds down to glassy states. The existence of such "serial decoupling" has been discussed in the literature, and even the need to recognize "second fictive temperatures" has been noted. It is interesting to realize, from the study by Zanotto and coworkers in these proceedings, that devitrification of complex liquids is usually controlled by a specific member of such a hierarchy.

**Acknowledgements.**
This work has been supported by the NSF-DMR under Solid State Chemistry Grant no. 0082535. The author has profited from extensive discussions with colleagues Dmitry Matyushov and Ranko Richert. We acknowledge also the support of the Brazilian Science Foundation for generous support for academic activities in Brazil during which many of the ideas contained in this article were assembled.

**References**

[1] M. H. Cohen and D. Turnbull, *Nature* **203** (1964), p. 964.
[2] M. H. Cohen and D. Turnbull, *J. Chem. Phys.* **34,** (1960).
[3] W. H. Wang, J. J. Lewandwski and L. Greer, *J. Mater. Res.* **20** (2005).
[4] K. Vollmayr, W. Kob and K. Binder, *J. Chem. Phys.* **105** (1996), p. 4714.
[5] E. Wachtel, I. Bakonyi, J. Bahle, N. Willmann, A. Lovas, A. Burgstaller, W. Socher, J. Voitlander and H. H. Liebermann, *Materials Science and Engineering A-Structural Materials Properties Microstructure and Processing* **133** (1991), p. 196.
[6] W. Kob, *Journal of Physics-Condensed Matter* **11** (1999), p. R85.
[7] B. G. Bagley and D. Turnbull, *J. Appl. Phys.* **39** (1968), p. 5681.
[8] T. A. Weber and F. H. Stillinger, *Phys. Rev B* **31** (1985), p. 1954.
[9] W. Kob and H. C. Andersen, *Phys. Rev. E* **51** (1995), p. 4626.
[10] P. Harrowell, ((private communication)).
[11] C. A. Angell, *Nature* **393** (1998), p. 521.
[12] R. Lynden-Bell and P. G. Debenedetti, *J. Phys. Chem. B* **109** (2005), p. 6527.
[13] V. Molinero, S. Sastry and C. A. Angell, *Physical Review Letters* **97** (2006).
[14] F. H. Stillinger and T. A. Weber, *Physical Review B* **31** (1985), p. 5262.
[15] F. Spaepen and D. Turnbull, *AIP Conference Proceedings,. Eds. Ferris S.D., Leamy, H. J., and Poate, J.* (1978), p. 73.
[16] B. G. Bagley and H. S. Chen, *AIP Conference Proceedings,. Eds. Ferris S.D., Leamy, H. J., and Poate, J.* (1978), p. 97.
[17] C. A. Angell and S. Borick, *J. Phys. Condens. Matter* **11** (1999), p. 8163.
[18] L. I. Aptekar, *Sov. Phys. Dokl.* **24** (1979), p. 993.
[19] I. L. Aptekar and E. G. Ponyatovsky, *Phys. Met. Metallogr.* **25** (1968), p. 10 and 93.
[20] S. M. Chathoth, A. Meyer, M. M. Koza and F. Juranyi, *Applied Physics Letters* **85** (2004), p. 4881.

[21] C. A. Angell, S. Borick and M. Grabow, *Journal of Non-Crystalline Solids* **207** (1996), p. 463.
[22] S. Sastry and C. A. Angell, *Nature Materials* **2** (2003), p. 739.
[23] W. B. Hillig and D. Turnbull, *J. Chem. Phys.* **24** (1956), p. 914.
[24] H. Bhat, V. Molinero, E. Soignard, V. C. Solomon, S. Sastry, J. L. Yarger and C. A. Angell, *Nature* **448 (7155), 787-U3** (2007), p. 787.
[25] B. J. Park, W. T. Kim, D. H. Kim and H. K., *MRS Fall meeting, Boston. Symposium on BMG, invited lecture Z6.6.*
[26] T. Ohkubo, Y. C. Kim, E. Fleury and K. Hono, *Scripta Materialia* **53** (2005), p. 165.
[27] S. Aasland and P. F. Mcmillan, *Nature* **369** (1994), p. 633.
[28] R. Kurita and H. Tanaka, *Phys. Rev. Lett.* **92** (2004), p. 025701
[29] W. A. Crichton, M. Mezouar, T. Grande, S. Stolen and A. Grzechnik, *Nature* **414** (2001), p. 622.
[30] E. A. P. R. Jagla, *Phys. Rev. E* **58** (1998), p. 1478.
[31] Z. Yan, S. V. Buldyrev, G. N., P. G. Debenedetti and H. E. Stanley, *Phys. Rev. E* **73** (2006), p. 051204.
[32] L. Xu, P. Kumar, S. V. Buldyrev, S.-H. Chen, P. H. Poole, F. Sciortino and H. E. Stanley, *Proc. Nat. Acad. Sci.* **102** (2005), p. 16558
[33] H. M. Gibson and N. B. Wilding, *Phys. Rev. E* **73** (2006), p. 061507.
[34] L.-M. Xu, S. V. Buldyrev, N. Giovambattista, C. A. Angell and H. E. Stanley, *(submitted)* (2007).
[35] A. Barros de Oliveira, P. A. Netz, T. Colla and M. C. Barbosa, *J. Chem. Phys.* **124** (2007 ), p. 84505.
[36] D. Turnbull and R. E. Cech, *J. Appl. Phys.* **21** (1950), p. 804.
[37] H. W. Sheng, H. Z. Liu, Y. Q. Cheng, J. Wen, P. L. Lee, W. K. S. Luo, S. D. and E. Ma, *Nature Materials* **6** (2007), p. 192.
[38] B. Zhang, R. J. Wang, D. Q. Zhao, M. X. Pan and W. H. Wang, *Phys. Rev B* **70** (2004), p. 224208s.
[39] S. V. Buldyrev, P. Kumar, P. G. Debenedetti, P. J. Rossky and H. E. Stanley, *Proc. Nat. Acad. Sci.* **online doi/10.1073/pnas.o708427104** (2007).
[40] C. A. Angell, *Science* **267** (1995), p. 1924.
[41] M. H. Cohen and G. S. Grest, *Annals of the New York Academy of Sciences* **371** (1981), p. 199.
[42] C. A. Angell, K. L. Ngai, G. B. McKenna, P. F. McMillan and S. W. Martin, *J. Appl. Phys.* **88** (2000), p. 3111.
[43] J. Dyre, *Revs. Mod. Phys.* **78** (2006), p. 953.
[44] I. Avramov, *J. Non-Cryst. Solids* **238** (1998), p. 6.
[45] L. M. Martinez and C. A. Angell, *Nature* **410** (2001), p. 663.
[46] G. Adam and J. H. Gibbs, *Journal of Chemical Physics* **43** (1965), p. 139.
[47] J. C. Dyre, N. B. Olsen and T. Christensen, *Phys. Rev. B* **53** (1996), p. 2171.
[48] D. Matyushov and C. A. Angell, *J. Chem. Phys.* **126** (2007), p. AN094501.
[49] M. M. Hurley and P. Harrowell, *Phys. Rev. E* **52** (1995), p. 1694.
[50] M. D. Ediger, *Ann. Rev. Phys. Chem.* **51** (2000), p. 99.
[51] R. Richert, *J. Phys. Condens. Matter* **14** (2002), p. R703.

[52] C. Donati, J. F. Douglas, W. Kob, P. H. Poole, S. J. Plimpton and S. C. Glotzer, *Phys. Rev. Lett.* **80** (1998), p. 2338.
[53] E. R. Weeks, J. C. Crocker, A. C. Levitt and D. A. Weitz, *Science* **287** (2000), p. 627.
[54] L. Lyon and L. T. A., *J. Appl. Phys.* **27** (1956), p. 129.
[55] C. A. Angell, *Annual Review of Physical Chemistry* **34** (1983), p. 593.
[56] L. M. Wang and R. Richert, *J. Chem. Phys.* **121** (2004), p. 11170.
[57] H. Huth, L.-M. Wang, C. Schick and R. Richert, *J. Chem. Phys.* **126** (2007), p. 104503.
[58] A. Widmer-Cooper and P. Harrowell, *J. Phys. Condens. Matter* **17** (2005), p. S4025.
[59] A. Widmer-Cooper and P. Harrowell, *J. Chem. Phys.* **126** (2007), p. 154403.
[60] R. V. Chamberlin, *NSF Research Proposal (s)* (1990).
[61] B. Schiener, R. Bohmer, A. Loidl and R. V. Chamberlin, *Science* **274** (1996), p. 752.
[62] R. Richert and S. Weinstein, *Phys. Rev. Lett.* **97** (2006), p. 095703.1
[63] L.-M. Wang and R. Richert, *Phys. Rev. Lett.* **99** (2007), p. 185701(1.
[64] M. Goldstein, *J. Chem. Phys.* **64** (1976), p. 4767.
[65] C. A. Angell, L.-M. Wang, S. Mossa and J. R. D. Copley, *A. I. P. Conference Proceedings ("Slow Dynamics" 2003)* **708** (2004), p. 473.
[66] S. Tamura, T. Yokokawa and K. J. Niwa, *J. Chem. Thermodyn.* **7** (1975), p. 633.
[67] S. Tamura and T. Yokokawa, *Bull. Chem. Soc. Japan* **48** (1975), p. 2542.
[68] M. Hemmati, C. T. Moynihan and C. A. Angell, *J. Chem. Phys.* **115** (2001), p. 6663.
[69] P. Scheidler, W. Kob, A. Latz, J. Horbach and K. Binder, *Phys. Rev. B* **63** (2005), p. 104204.
[70] C. A. Angell, *Science (in press)* (2008).
[71] C. A. Angell, *Chem. Rev.* **102,** (2002), p. 2627.
[72] R. Kurita and H. Tanaka, *Science* **306** (2004), p. 845.
[73] R. Kurita and H. Tanaka, *J. Phys. Conden. Matter* **17** (2005), p. L293.
[74] C. A. Angell, *Pergammon Encyclopedia of Materials: Science and Technology, 3365* **4** (2001), p. 3365.
[75] K. Adachi, H. Suga and S. Seki, *Bull. Chem. Soc. Japan* **41** (1968), p. 1073.
[76] G. P. Johari, *Ann. N. Y. Acad. Sci.,* **279** (1976), p. 117.
[77] O. Haida, H. Suga and S. Seki, *J. Chem. Thermodynamics* **9** (1977), p. 1133.
[78] C. A. Angell, *J. Non-Cryst. Solids* **131-133** (1991), p. 13.
[79] R. Brand, A. Loidl and P. Lunkenheimer, *J. Chem. Physics* **116** (2002), p. 10386.
[80] W. Kauzmann, *Chem. Rev.* **43** (1948), p. 218.
[81] See notes in Wikipedia under "order-disorder transitions" and "lambda transitions"
[82] D. Wales, J., M. A. Miller and T. R. Walsh, *Nature* **394** (1998), p. 758.
[83] D. J. Wales, *Energy Landscapes* **Cambridge Molecular Science** (2003), p. 654pp.
[84] T. Matsuo, H. Suga, W. I. F. David, R. M. Ibberson, P. Bernier, A. Zahab, C. Fabre, A. Rassat and A. Dworkin, *Solid State Communications* **83** (1992), p. 711.

[85] P. Mondal, P. Lunkenheimer and A. Loidl, *Zeitschrift Fur Physik B-Condensed Matter* **99** (1996), p. 527.
[86] S. Kaya and H. Sato, *Proc. PhysicoMath. Soc. Japan* **25** (1943), p. 261.
[87] L.-M. Wang and C. A. Angell, *J. Non-Cryst. Solids* **353 (41-43)** (2007), p. 3829.
[88] Y. Z. Yue, S. L. Jensen and J. D. Christiansen, *Applied Physics Letters* **81** (2002), p. 2983.
[89] R. Bohmer and C. A. Angell, *Phys. Rev B* **45** (1992), p. 10091.
[90] M. H. J. Hagen, E. J. Meijer, G. C. A. M. Moolj, D. Fenkel and H. N. W. Lekkerkerker, *Nature* **365** (1993), p. 425.